\documentclass[pra,10pt,twocolumn,aps,showpacs,superscriptaddress]{revtex4-1}

\usepackage{amsmath,amssymb,amsmath}
\usepackage{graphicx}
\usepackage{dcolumn}
\usepackage{bm}
\usepackage{multirow}
\usepackage[ansinew]{inputenc} 

\begin{document}

\title{Dressing of Ultracold Atoms by their Rydberg States in a Ioffe-Pritchard Trap}

\author{Michael Mayle}
\affiliation{Zentrum f\"ur Optische Quantentechnologien, Universit\"at
Hamburg, Luruper Chaussee 149, 22761 Hamburg, Germany}

\author{Igor Lesanovsky}
\affiliation{Midlands Ultracold Atom Research Centre - MUARC, The University of Nottingham,
School of Physics and Astronomy, Nottingham, United Kingdom}

\author{Peter Schmelcher}
\affiliation{Zentrum f\"ur Optische Quantentechnologien, Universit\"at
Hamburg, Luruper Chaussee 149, 22761 Hamburg, Germany}

\date{\today}

\begin{abstract}
We explore how the extraordinary properties of Rydberg atoms can be employed to impact the motion of ultracold ground state atoms. Specifically, we use an off-resonant two-photon laser dressing to map features of the Rydberg states on ground state atoms. It is demonstrated that the interplay between the spatially varying quantization axis of the considered Ioffe-Pritchard field and the fixed polarizations of the laser transitions provides the possibility of substantially manipulating the ground state trapping potential.
\end{abstract}


\maketitle

\section{Introduction}
Rydberg atoms are -- amongst others -- highly susceptible to external fields and show a strong mutual interaction \cite{gallagher94}. Combining the extraordinary properties of Rydberg atoms, which originate from the large displacement of the valence electron and the remaining ionic core, with the plethora of techniques known from the preparation and manipulation of ultracold gases enables remarkable observations such as the excitation blockade between two single atoms a few $\mu$m apart \cite{Urban2009,Gaetan2009}. Moreover, interacting Rydberg atoms serve as flexible tools for various purposes. For example, their strong dipole-dipole interaction renders Rydberg atoms interesting candidates for the realization of two-qubit quantum gates \cite{PhysRevLett.85.2208,PhysRevLett.87.037901} or efficient multiparticle entanglement \cite{saffman:240502,muller:170502,PhysRevLett.103.185302}. In fact, only very recently a {\sc cnot} gate between two individually addressed neutral atoms and the generation of entanglement has been demonstrated experimentally by employing the Rydberg blockade mechanism \cite{isenhower10,wilk10}.

Other proposals utilize the peculiar properties of an ensemble of interacting Rydberg atoms by employing an off-resonant laser coupling that dresses ground state atoms with Rydberg states. For example, a method has been proposed of creating a polarized atomic dipolar gas by coupling to an electrically polarized Rydberg state \cite{santos00}. The resulting long-ranged dipole-dipole interaction in such gases are predicted to give rise to dipolar crystals and novel supersolid phases \cite{pupillo10,cinti10}. In a similar manner, the Rydberg dressing of ground state atoms is expected to entail a roton-maxon excitation spectrum in three-dimensional Bose-Einstein condensates \cite{pohl10} and collective many-body interactions \cite{honer10}. Here, we discuss a further application for the use of Rydberg states, namely, how they can be employed for substantially manipulating the trapping potentials of magnetically trapped $^{87}$Rb atoms in a controlled manner.

Inhomogeneous magnetic trapping fields are omnipresent in experiments dealing with ultracold atoms. Constituting a promising alternative to optical approaches, even one- and two-dimensional lattices of magnetic microtraps have been realized experimentally \cite{guenter05,singh08,gerritsma07,whitlock09}.
The issue of trapping Rydberg atoms in magnetic traps -- primarily of Ioffe-Pritchard kind -- has been studied extensively, demonstrating that Rydberg atoms can be tightly confined \cite{hezel:223001,hezel:053417} and that one-dimensional Rydberg gases can be created and stabilized by means of an additional electric field \cite{mayle:113004}. In particular, the authors demonstrated in a previous work that the  trapping potentials of $^{87}$Rb Rydberg atoms in low angular momentum
electronic states (i.e., $l\leq2$) considerably deviate from the  behavior known from ground state atoms \cite{Mayle2009}. This effect is due to the \emph{composite} nature of Rydberg atoms, i.e., the fact that they consist of an outer valence electron far apart from the ionic core. 

In the present work we demonstrate how the peculiar properties of the Rydberg trapping potential can be utilized to manipulate the trapping potential for the ground state. To this end, an off-resonant two-photon laser transition is employed that dresses the ground state atoms by their Rydberg states. We thoroughly discuss the coupling scheme previously employed in \cite{mayle:041403} and systematically study the resulting dressed potentials. In particular, it is demonstrated how the delicate interplay between the spatially varying quantization axis of the Ioffe-Pritchard field and the fixed polarizations of the laser transitions greatly influences the actual shape of the trapping potentials -- a mechanism that has been been employed also very recently to create versatile atom traps by means of a Raman type setup \cite{middelkamp10}. Moreover, the employed scheme allows us to map the Rydberg trapping potential onto the ground state. 

In detail, we proceed as follows. Section \ref{sec:rydsurf} briefly reviews the properties of Rydberg atoms in a magnetic Ioffe-Pritchard trap as derived in \cite{Mayle2009}; the resulting trapping potentials are contrasted with the ones belonging to ground state atoms. Section \ref{sec:scheme} then introduces the off-resonant two-photon laser coupling scheme that dresses the ground state with the Rydberg state. In section \ref{sec:elimination} we establish a simplified three-level scheme (opposed to the 32-level scheme that is needed to fully describe the excitation dynamics) that allows us to derive analytical expressions of the dressed potentials. Section \ref{sec:dressedsurf} finally contains a thorough discussion of the dressed ground trapping potentials for a variety of field and laser configurations.

\section{Review of the Rydberg Trapping Potentials}
\label{sec:rydsurf}
Let us start by briefly recapitulating the results from \cite{Mayle2009} concerning the trapping potentials of alkali Rydberg atoms in their $nS$, $nP$, and $nD$ electronic states. As the basic ingredient for magnetically trapping Rydberg atoms, we consider the Ioffe-Pritchard field configuration given by $\mathbf{B}(\mathbf{x})=\mathbf B_c+\mathbf{B}_l(\mathbf{x})$ with $\mathbf B_c=B\mathbf e_3$, $\mathbf{B}_l(\mathbf{x})= G\left[x_1\mathbf{e}_1-x_2\mathbf{e}_2\right]$. The corresponding vector potential reads $\mathbf{A}(\mathbf{x})= \mathbf{A}_c(\mathbf{x})+\mathbf{A}_l(\mathbf{x})$, with $\mathbf{A}_c(\mathbf{x})= \frac{B}{2}\left[x_1\mathbf{e}_2-x_2\mathbf{e}_1\right]$ and $\mathbf{A}_l(\mathbf{x})=Gx_1x_2\mathbf{e}_3$; $B$ and $G$ are the Ioffe field strength and the gradient, respectively. The mutual interaction of the highly excited valence electron  and the remaining closed-shell ionic core of a Rydberg atom  is modeled by an effective potential which depends only on the distance of the two particles. After introducing relative and center of mass coordinates ($\mathbf{r}$ and $\mathbf{R}$) and employing the unitary transformation $U=\exp\left[\frac{i}{2}(\mathbf{B}_c\times \mathbf{r}) \cdot \mathbf{R}\right]$, the Hamiltonian describing the Rydberg atom becomes (atomic units are used unless stated otherwise)
\begin{eqnarray}
\label{eq:hamfinaluni}
H&=&H_A+\frac{\mathbf{P}^2}{2M}
+\frac{1}{2}[\mathbf L+2\mathbf S]\cdot\mathbf B_c
+\mathbf S\cdot\mathbf{B}_l(\mathbf{R+r})
\nonumber\\
&&+\mathbf{A}_l(\mathbf{R+r})\cdot\mathbf{p}
+H_\mathrm{corr}\,.
\end{eqnarray}
Here, $H_A=\mathbf{p}^2/2+V_l(r)+V_{so}(\mathbf{L},\mathbf{S})$ is the field-free Hamiltonian of the valence electron whose core penetration, scattering, and polarization effects are accounted for by the $l$-dependent model potential $V_l(r)$ \cite{PhysRevA.49.982} while $\mathbf L$ and $\mathbf S$ denote its orbital angular momentum and spin, respectively. $V_{so}(\mathbf{L},\mathbf{S})=\frac{\alpha^2}{2}\left[1-\frac{\alpha^2}{2}V_l(r)\right]^{-2}
\frac{1}{r}\frac{\mathrm d V_l(r)}{\mathrm d r}
\mathbf L\cdot\mathbf S$ denotes the spin-orbit interaction that couples $\mathbf L$ and $\mathbf S$ to the total electronic angular momentum $\mathbf J = \mathbf L + \mathbf S$; the term $\left[1-\alpha^2V_l(r)/2\right]^{-2}$ has been introduced to regularize the nonphysical divergence near the origin \cite{condon35}.
$H_\mathrm{corr}=-\boldsymbol{\mu}_c\cdot \mathbf{B(R)}+\frac{1}{2}\mathbf A_c(\mathbf r)^2+
\frac{1}{2}\mathbf A_l(\mathbf{R+r})^2
+\frac{1}{M}\mathbf B_c\cdot(\mathbf{r\times P})
+U^\dagger[V_l(r)+V_{so}(\mathbf{L},\mathbf{S})]U
-V_l(r)-V_{so}(\mathbf{L},\mathbf{S})$ are small corrections that are neglected in the  parameter regime we are focusing on; the magnetic moment of the ionic core is connected to the nuclear spin $\mathbf I$ according to $\boldsymbol{\mu}_c=-\frac{1}{2}g_I\mathbf I$, with $g_I$ being the nuclear g-factor. In order to solve the resulting coupled Schr\"odinger equation, we employ a Born-Oppenheimer separation of the center of mass motion and the electronic degrees of freedom. We are thereby led to an electronic Hamiltonian for fixed center of mass position of the atom whose eigenvalues $E_\kappa(\mathbf R)$ depend parametrically on the center of mass coordinates. These adiabatic electronic surfaces serve as trapping potentials for the quantized center of mass motion.

For fixed total electronic angular momentum $\mathbf{J=L}+\mathbf S$, approximate expressions for the adiabatic electronic energy surfaces can be derived by applying the spatially dependent transformation $U_r=e^{-i \gamma (L_x+S_x)}e^{-i \beta (L_y+S_y)}$ that rotates the local magnetic field vector into the $z$-direction of the laboratory frame of reference. The corresponding rotation angles are defined by
$\sin\gamma=-GY/\sqrt{B^2+G^2(X^2+Y^2)}$,
$\sin\beta=-GX/\sqrt{B^2+G^2X^2}$,
$\cos\gamma =\sqrt{B^2+G^2X^2}/\sqrt{B^2+G^2(X^2+Y^2)}$, and
$\cos\beta =B/\sqrt{B^2+G^2X^2}$.
In second order perturbation theory, the adiabatic electronic energy surfaces read
\begin{equation}\label{eq:dressealpha}
 E_\kappa(\mathbf R)= E_\kappa^{(0)}(\mathbf R)+E_\kappa^{(2)}(\mathbf R),
\end{equation}
where
\begin{equation}\label{eq:ealpha0}
E_\kappa^{(0)}(\mathbf R) = E_\kappa^{el}+\frac{1}{2}g_jm_j\sqrt{B^2+G^2(X^2+Y^2)}
\end{equation}
represents the coupling of a point-like particle to the magnetic field via its magnetic moment $\boldsymbol{\mu}\propto\mathbf J=\mathbf L+\mathbf S$; $\kappa$ represents the electronic state under investigation, i.e., $|\kappa\rangle=|njm_jls\rangle$, $g_j=\frac{3}{2}+\frac{s(s+1)-l(l+1)}{2j(j+1)}$ its Land\'e g-factor, and $E_\kappa^{el}$ the field-free atomic energy levels. $E_\kappa^{(0)}(\mathbf R)$ is rotationally symmetric around the $Z$-axis  and confining for $m_j>0$. For small radii ($\rho=\sqrt{X^2+Y^2}\ll B/G$) an expansion up to second order yields a harmonic potential
\begin{equation}
E_\kappa^{(0)}(\rho)\approx E_\kappa^{el}+\frac{1}{2}g_jm_jB+
\frac{1}{2}M\omega^2\rho^2
\end{equation}
with the trap frequency defined by $\omega=G\sqrt{\frac{g_jm_j}{2MB}}$ while we find a linear behavior $E_\kappa^{(0)}(\rho)\approx E_\kappa^{el}+\frac{1}{2}g_jm_jG\rho$ when the center of mass is far from the $Z$-axis ($\rho\gg B/G$).

The second order contribution $E_\kappa^{(2)}(\mathbf R)$ stems from the composite nature of the Rydberg atom, i.e., the fact that it consists of an outer Rydberg electron far apart from the ionic core. It reads 
\begin{equation}\label{eq:ealpha2}
 E_\kappa^{(2)}(\mathbf R)=C G^2X^2Y^2\,,
\end{equation}
where the coefficient $C$ depends on the electronic state $\kappa$ under investigation. Since $C$ is generally negative \cite{Mayle2009}, a de-confining behavior of the energy surface for large center of mass coordinates close to the diagonal ($X=Y$) is found. For a detailed derivation and discussion of the Rydberg trapping
potentials (\ref{eq:dressealpha}-\ref{eq:ealpha2}) we refer the reader to \cite{Mayle2009}.

\subsection*{Trapping Potentials of Ground State Atoms}
When considering the trapping of ground state atoms, the coupling mechanism relies on the point-like interaction of the atomic magnetic moment $\boldsymbol{\mu}$ with the external field. Since the hyperfine interaction easily overcomes the Zeeman splitting for the regime of magnetic field strengths we are interested in, we include the hyperfine interaction in our theoretical considerations and assume the atom to couple via its total angular momentum $\boldsymbol{\mu}\propto\mathbf F=\mathbf J+\mathbf I$ to the magnetic field ($\mathbf I$ being the nuclear spin). The ground state trapping potentials correspondingly read
\begin{equation}
 E_\kappa(\mathbf R)=E_\kappa^{el}+\frac{1}{2}g_Fm_F|\mathbf{B(R)}|\,,
\end{equation}
where $E_\kappa^{el}$ includes the hyperfine as well as spin-orbit effects, and
\begin{equation}
 g_F=g_j\frac{F(F+1)+j(j+1)-I(I+1)}{2F(F+1)}\,.
\end{equation}
Let us note that for Rydberg atoms the hyperfine interaction $H_\mathrm{hfs}=A\mathbf I\cdot\mathbf J$ only plays a minor role since the hyperfine constant $A$ scales as $n^{-3}$ \cite{li:052502}. For a wide range of field strengths it is therefore sufficient to treat the hyperfine interaction perturbatively, giving rise to a mere splitting of the Rydberg trapping potentials (\ref{eq:dressealpha}) according to $W_\mathrm{hfs}=Am_im_j$ \cite{armstrong71}. Correspondingly, we continue to label the Rydberg states by their $j$, $m_j$, and $m_I$ quantum numbers rather than the $F$, $m_F$ ones. In particular, for characterizing the Rydberg trapping potentials the $j$, $m_j$ quantum numbers are sufficient. In our numerical calculations, on the other hand, we fully incorporated the hyperfine interaction $H_\mathrm{hfs}$ of the Rydberg state. Moreover, we also included the coupling of the magnetic moment of the ionic core, $\boldsymbol{\mu}\propto\mathbf I$, to the field.
Finally, we remark that -- except for the electronic energy offset $E_\kappa^{el}$ -- the zeroth order Rydberg trapping potential $E_{nS_{1/2}, m_j=1/2}^{(0)}(\mathbf R)$ and the $5S_{1/2}$ ground state energy surface are identical for $F=m_F=2$.

\section{Off-Resonant Coupling Scheme}
\label{sec:scheme}
In this section, we discuss the coupling scheme of the ground- and Rydberg state that arises for a two-photon off-resonant laser excitation in the presence of the Ioffe-Pritchard trap. The off-resonant coupling results in a dressed ground state atom to which the Rydberg state is weakly admixed. In this manner, the ground state atom gains properties that are specific for the Rydberg atom. In particular, the peculiar properties of the Rydberg trapping surfaces can be exploited for substantially manipulating the trapping potentials of ground state atoms.

\begin{figure}
\centering
\includegraphics[width=8.6cm]{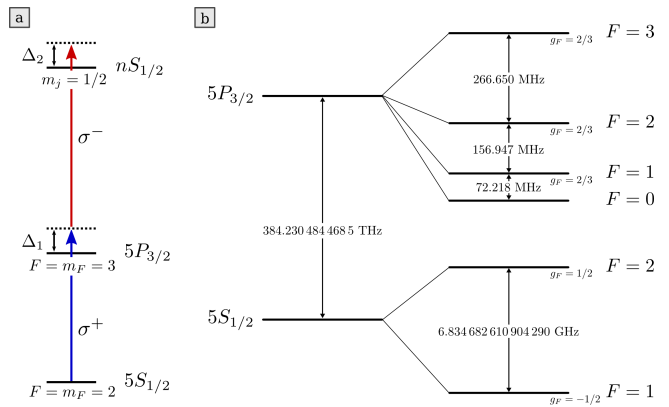}
\caption{(a) Idealized level scheme for an off-resonant two photon coupling of the ground- and Rydberg state of $^{87}$Rb. In a Ioffe-Pritchard trap, additional atomic levels and polarizations contribute away from the trap center, see text. Note that the hyperfine splittings of the Rydberg level are included in the calculation although not shown in this figure. (b) Atomic energy level scheme of the $5S_{1/2}$ and $5P_{3/2}$ states of $^{87}$Rb including the hyperfine splittings.}
\label{fig:dressscheme} 
\end{figure}

We investigate the excitation scheme that is frequently encountered in experiments \cite{heidemann:163601,reetz-lamour:253001}: Laser 1, which is $\sigma^+$ polarized, drives the transition $s\rightarrow p$ detuned by $\Delta_1$ while a second, $\sigma^-$ polarized laser then couples to the Rydberg state  $n\equiv nS_{1/2},m_j=1/2, m_I=3/2$, with $s$ denoting the ground state $5S_{1/2}, F=m_F=2$ and $p$ the intermediate state $5P_{3/2},F=m_F=3$. Both lasers are propagating along the $\mathbf{e}_3$-axis in the laboratory frame of reference; the complete two-photon transition is supposed to be off-resonant by $\Delta_2$. A sketch of the whole scheme is provided in figure \ref{fig:dressscheme}(a). In a Ioffe-Pritchard trap, however, the quantization axis is spatially dependent and the polarization vectors of the two excitation lasers are only well defined as $\sigma^+$ and $\sigma^-$ at the trap center. As we are going to show in the following, in the rotated frame of reference, i.e., after applying the unitary transformation $U_r$, contributions of all polarizations emerge and the excitation scheme becomes more involved.

In the dipole approximation, the interaction of the atom with the laser fields is given by
\begin{equation}
H_\mathrm{AF}=-\sum_{i=1}^{2}\mathbf{d\cdot E}_i(t)=\sum_{i=1}^{2}\mathbf{r\cdot E}_i(t)\,,
\label{eq:hl}
\end{equation}
where the sum runs over the two applied excitation lasers. The electric field vectors $\mathbf E_i(t)$ can be decomposed into their positive- and negative-rotating components according to $\mathbf E_i^{(+)}(t)$ and $\mathbf E_i^{(-)}(t)$,
\begin{eqnarray}
  \mathbf E_i(t)&=&\frac{E_{i0}}{2}\big(\boldsymbol{\epsilon}_ie^{-i\omega t}+\boldsymbol{\epsilon}_i^*e^{i\omega t}\big)\\
&\equiv& \mathbf E_i^{(+)}(t)+\mathbf E_i^{(-)}(t)\,,
\end{eqnarray}
i.e., $\mathbf E_i^{(\pm)}\propto e^{-i(\pm \omega)t}$. The electric field amplitude $E_{i0}$ is connected to the intensity $I_i$ of the $i$th laser via $E_{i0}=\sqrt{2 I_i/c\varepsilon_0}$. We distinguish three different polarization vectors $\boldsymbol{\epsilon}$ of the excitation lasers, namely, 
{$\boldsymbol{\epsilon}_\pm=(\mathbf e_1\pm i\mathbf e_2)/\sqrt{2}$ and $\boldsymbol{\epsilon}_0=\mathbf e_3$ for $\sigma^\pm$- and $\pi$-polarized light, respectively.

In order to solve the time-dependent Schr\"odinger equation, the Hamiltonian for the atom in the Ioffe-Pritchard trap and the laser interaction must be expressed in the same frame of reference. Hence, the unitary transformations of the previous section must be applied to $H_\mathrm{AF}$ as well. The first one, $U=\exp\left\{\frac{i}{2}(\mathbf{B}_c\times \mathbf{r}) \cdot \mathbf{R}\right\}$, leaves the interaction Hamiltonian (\ref{eq:hl}) of the atom with the lasers unchanged. The transformation $U_r=e^{-i \gamma J_x}e^{-i \beta J_y}$ into the rotated frame of reference, on the other hand, yields 
\begin{equation}
 U_r\mathbf rU_r^\dagger=
\left(\begin{array}{c}
x\cos\beta + y\sin\gamma\sin\beta - z\cos\gamma\sin\beta\\
y\cos\gamma + z\sin\gamma\\
x\sin\beta - y\sin\gamma\cos\beta + z\cos\gamma\cos\beta
\end{array}\right)\,.
\end{equation}
That is, the $\sigma^+$ and $\sigma^-$ laser transitions that are depicted in figure \ref{fig:dressscheme}(a) become
\begin{eqnarray}
 \boldsymbol{\epsilon}_\pm\cdot U_r\mathbf rU_r^\dagger&=&\frac{1}{\sqrt{2}}\big[x\cos\beta+y\sin\gamma\sin\beta\nonumber\\
&&\quad\,\,-z\cos\gamma\sin\beta
\pm i(y\cos\gamma+z\sin\gamma)\big]\,.\label{eq:dressUeps}
\end{eqnarray}
Equation (\ref{eq:dressUeps}) can be rewritten in terms of the polarization vectors $\tilde{\boldsymbol{\epsilon}}_\pm$ and $\tilde{\boldsymbol{\epsilon}}_0$ defined in the rotated frame of reference. To this end, we rotate the polarization vector $\boldsymbol{\epsilon}$ and leave the position operator $\mathbf r$ unchanged:
$\boldsymbol{\epsilon}\cdot U_r\mathbf rU_r^\dagger\rightarrow(\mathcal R\boldsymbol{\epsilon})\cdot\mathbf r$ with $\mathcal R$ denoting the rotation matrix associated with the transformation $U_r$. $\mathcal R\boldsymbol{\epsilon}$ can then be decomposed into the components $\tilde{\boldsymbol{\epsilon}}_\pm$ and $\tilde{\boldsymbol{\epsilon}}_0$, i.e., $\mathcal R\boldsymbol{\epsilon}=\sum_{i=\pm,0} c_i\tilde{\boldsymbol{\epsilon}_i}$ with $c_i=\tilde{\boldsymbol{\epsilon}}_i^*\cdot\mathcal R\boldsymbol{\epsilon}$. Employing 
\begin{equation}
 \mathcal R\boldsymbol{\epsilon}_\pm=\frac{1}{\sqrt{2}}
\left(\begin{array}{c}
 \cos\beta\\
\sin\gamma\sin\beta\pm i\cos\gamma\\
-\cos\gamma\sin\beta\pm i\sin\gamma
\end{array}\right)
\end{equation}
finally yields
\begin{eqnarray}
 \boldsymbol{\epsilon}_+\cdot U_r\mathbf rU_r^\dagger&=&
\bigg[\frac{1}{2}(\cos\gamma+\cos\beta-i\sin\gamma\sin\beta)\tilde{\boldsymbol{\epsilon}}_+\nonumber\\
&&\,\,-\frac{1}{2}(\cos\gamma-\cos\beta-i\sin\gamma\sin\beta)\tilde{\boldsymbol{\epsilon}}_-\nonumber\\
&&\,\,-\frac{1}{\sqrt{2}}(\cos\gamma\sin\beta-i\sin\gamma)\tilde{\boldsymbol{\epsilon}}_0\Big]\cdot\mathbf r\,,\label{eq:dress_epsplus}\\
\boldsymbol{\epsilon}_-\cdot U_r\mathbf rU_r^\dagger&=&(\boldsymbol{\epsilon}_+\cdot U_r\mathbf rU_r^\dagger)^*.\label{eq:dress_epsminus}
\end{eqnarray}
Thus, in the rotated frame of reference contributions of all polarizations emerge away from the trap center. In particular, the $5S_{1/2},F=m_F=2$ ground state can also couple to $m_F<3$ magnetic sublevels of the $5P_{3/2}$ intermediate state. Moreover, two-photon couplings between the $5S_{1/2},F=m_F=2$ and $5S_{1/2},F=2, m_F<2$ levels via the hyperfine levels of the $5P_{3/2}$ intermediate state emerge if the first excitation laser gains a significant contribution of the $\sigma^-$- or $\pi$-polarization in the rotated frame of reference. On the Rydberg side, also $m_j=-1/2$ states become accessible. As a results, the simple three-level excitation scheme $s \leftrightarrow p \leftrightarrow n$ is in general not sufficient and all relevant hyperfine levels must be included in the theoretical treatment. 
In detail, these are the $F=1$ and $F=2$ hyperfine levels of the $5S_{1/2}$ ground and the $nS_{1/2}$ Rydberg state. For the intermediate $5P_{3/2}$ state we have $F\in\{0,1,2,3\}$. Of course, for each $F$ there are in addition $2F+1$ magnetic sublevels with $|m_F|\le F$. Note that the intermediate $5P_{3/2},F<3$ states are split considerably below the $5P_{3/2},F=3$ levels because of the hyperfine interaction, see figure \ref{fig:dressscheme}(b). An even stronger splitting is encountered for the $F=1$ and $F=2$ hyperfine levels of the $5S_{1/2}$ electronic state. Nevertheless, all these states are taken into account in our numerical calculations, yielding in total 32 states. Other electronic states are far off-resonant and thus do not contribute in the excitation dynamics.

The resulting multi-level excitation scheme is solved by employing the rotating wave approximation while adiabatically eliminating the intermediate states by a strong off-resonance condition. This procedure results in an effective coupling matrix for the ground and Rydberg states whose diagonalization yields a dressed electronic potential energy surface for the center of mass motion of the ground state atom. In the next section, we derive the coupling matrix for the illustrative example of a simplified three-level system. The generalization to the full level scheme is straightforward, although laborious.

\section{Simplified Three-Level Scheme}
\label{sec:elimination}
In this section, we restrict ourselves to the three-level system $s\leftrightarrow p\leftrightarrow n$, i.e., including from the transformed dipole interaction (\ref{eq:dress_epsplus}-\ref{eq:dress_epsminus}) only the $\sigma^+$ and $\sigma^-$ part for the first and second laser, respectively. Such a simplification allows us to derive analytical solutions of the time-dependent Schr\"odinger equation and therefore constitutes a particularly illustrative example. It is expected to be valid for large Ioffe fields $B$ and/or small gradients $G$ when the quantization axis only shows a weak spatial dependence and  the $nS_{1/2},m_j=1/2,m_I=3/2$ Rydberg state is predominantly addressed via the $5P_{3/2},F=m_F=3$ intermediate state. For higher gradients, the polarization vector significantly changes its character throughout the excitation area such that the contributions of other states cannot be neglected anymore.

The three-level system can be further simplified by adiabatically eliminating the intermediate state $p$ by a strong off-resonance condition, i.e., assuming $|\Delta_1|\gg\omega_{ps}$ and $|\Delta_1-\Delta_2|\gg\omega_{np}$ 
with $\omega_{ps}$ and $\omega_{np}$ being the single-photon Rabi frequencies of the first and second laser transition, respectively:
\begin{eqnarray}
 \omega_{ps}&=&\frac{1}{2}(\cos\gamma+\cos\beta-i\sin\gamma\sin\beta)\cdot\omega_{ps}^{(0)}=\omega_{sp}^*\label{eq:rabi1}\,,\\
 \omega_{np}&=&\frac{1}{2}(\cos\gamma+\cos\beta+i\sin\gamma\sin\beta)\cdot\omega_{np}^{(0)}=\omega_{pn}^*\,.\label{eq:rabi2}
\end{eqnarray}
$\omega_{ps}^{(0)}=E_{1,0}\langle p|\tilde{\boldsymbol{\epsilon}}_+\cdot\mathbf r|s\rangle$ and $\omega_{np}^{(0)}=E_{2,0}\langle n|\tilde{\boldsymbol{\epsilon}}_-\cdot\mathbf r|p\rangle$ denote the single-photon Rabi frequency at the trap center. We remark that in the regime of strong Ioffe fields, where the simplified three-level scheme is valid, the spatial dependencies of (\ref{eq:rabi1}-\ref{eq:rabi2}) are largely negligible. Hence, the single-photon Rabi frequencies are to a good approximation given by their values $\omega_{ps}^{(0)}$ and $\omega_{np}^{(0)}$ at the origin. 
Employing in addition the rotating wave approximation, quasidegenerate van Vleck perturbation theory \cite{shavitt:5711} provides us an effective two-level Hamiltonian for the ground state $s$ and the Rydberg state $n$:
\begin{equation}
 \mathcal H_{2l}=
\left(\begin{array}{ccc}
 \Delta_2+\tilde{E}_n+E_\mathrm{hfs}+V_n& \Omega/2\\
 \Omega^*/2&\tilde{E}_s+V_s
\end{array}\right).
\label{eq:h2l}
\end{equation}
Here, $E_\mathrm{hfs}$ includes the energy shift due to the hyperfine splitting of the Rydberg state as well as the Zeeman shift of the nuclear spin.
For a detailed derivation of Hamiltonian (\ref{eq:h2l}) we refer the reader to the appendix of this work. 
$\tilde E_n\equiv \frac{1}{2}|\mathbf{B(R)}|+C\cdot G^2X^2Y^2$,
$\tilde E_s\equiv \frac{1}{2}|\mathbf{B(R)}|$, and
$\tilde E_p\equiv |\mathbf{B(R)}|$
are the trapping potentials of the individual energy levels. Note that the Rydberg state $n$ experiences the same potential energy surface as the ground state $s$, plus the perturbation $E_{nS_{1/2}}^{(2)}(\mathbf R)$ due to its non-pointlike character. The laser detunings are defined by $\Delta_1=E_p^{el}-E_s^{el}-\omega_1$ and $\Delta_2=E_n^{el}-E_s^{el}-\omega_1-\omega_2$.
The effective interaction between the ground- and Rydberg state is given by the two-photon Rabi frequency 
\begin{equation}
 \Omega=\frac{\omega_{ps}\omega_{np}}{4}
\left[\frac{1}{\tilde{E}_s-\tilde{E}_p-\Delta_1}+\frac{1}{\tilde{E}_n-\tilde{E}_p+\Delta_2-\Delta_1+E_\mathrm{hfs}}\right]\,.
\label{eq:omega}
\end{equation}
On the diagonal of Hamiltonian (\ref{eq:h2l}) we find the contributions
\begin{eqnarray}
 V_n&=-\frac{1}{4}\frac{|\omega_{np}|^2}{\tilde{E}_p-\tilde{E}_n+\Delta_1-\Delta_2-E_\mathrm{hfs}}\,,
\label{eq:vn}\\
 V_s&=-\frac{1}{4}\frac{|\omega_{ps}|^2}{\tilde{E}_p-\tilde{E}_s+\Delta_1}\,,
\label{eq:vs}
\end{eqnarray}
which are the light shifts of the Rydberg and ground state, respectively. In the limit $\Delta_1\gg\Delta_2$ and neglecting the energy surfaces $\tilde{E}_i$ -- which means looking at the trap center -- one recovers $\Omega=-\omega_{ps}\omega_{np}/2\Delta_1$, $V_n=-|\omega_{np}|^2/4\Delta_1$, and $V_s=-|\omega_{ps}|^2/4\Delta_1$. 

The diagonalization of Hamiltonian (\ref{eq:h2l}) yields the \emph{dressed} Rydberg ($+$) and ground state energy surfaces ($-$), 
\begin{eqnarray}
 E_\pm(\mathbf R)&=&\frac{1}{2}\bigg[\tilde{E}_s+V_s+\tilde{E}_n+V_n+\Delta_2+E_\mathrm{hfs}\nonumber\\
&&\quad\pm\sqrt{(\tilde{E}_n+V_n-\tilde{E}_s-V_s+\Delta_2+E_\mathrm{hfs})^2+\Omega^2}\bigg],\label{eq:dressed}
\end{eqnarray}
that serve as trapping potential for the external motion. Here, we are mainly interested in the dressed potential for the ground state. For large detunings $\Delta_2\gg\Omega$ one can approximate
\begin{equation}\label{eq:dressed2}
 E_-(\mathbf R)\approx\tilde{E}_s+V_s-\frac{\Omega^2}{4\Delta_2}+\frac{\Omega^2}{4\Delta_2^2}(\tilde{E}_n+V_n-\tilde{E}_s-V_s)\,,
\end{equation}
i.e., the contribution of the Rydberg surface $\tilde E_n$ to the dressed ground state trapping potential $E_-(\mathbf R)$ is suppressed by the factor $(\Omega/\Delta_2)^2$. Note that any spatial variation in the light shift $V_s$ and in the Rabi frequency $\Omega$ will effectively alter the trapping potential experienced by the dressed ground state atom.

\section{Dressed Ground State Trapping Potentials}
\label{sec:dressedsurf}
In this section, we investigate the dressed ground state trapping potentials arising from the two-photon coupling described in Section \ref{sec:scheme}. Since the actual shape of these energy surfaces is determined by the interplay of the various parameters belonging to the field configuration ($B$ and $G$) as well as to the laser couplings ($\omega_{ps}^{(0)}$, $\omega_{np}^{(0)}$, $\Delta_1$, and $\Delta_2$), there is a plethora of possible configurations. Nevertheless, one can distinguish basically two relevant regimes based on the magnetic field parameters. First of all, there is the regime where the ground state trapping potential is substantially influenced by the admixture of the Rydberg surface. This regime is usually encountered for a Ioffe dominated magnetic field configuration combined with a relatively strong laser coupling. In contrast, the second regime is obtained for strong gradient fields. In this case, the resulting spatially inhomogeneous light shift determines the characteristics of the ground state trapping potential and the contribution of the Rydberg surface is of minor importance. Exemplary dressed energy surfaces belonging to both regimes are discussed in the following. We stress that for determining the dressed trapping potentials the full 32-level scheme is solved. Comparisons with the analytically obtained result (\ref{eq:dressed}) are provided. Concerning the choice of the Rydberg state $n$, a principal quantum number of $n=40$ is considered throughout this section.

\subsection{Dressed Trapping Potentials of the $m_F=2$ State}
Let us start by investigating the dressed potential arising for the $5S_{1/2},\,m_F=2$ state of the rubidium atom. As mentioned before, in zero order this state gives rise to the same trapping potential as the $nS_{1/2}$ Rydberg state. Hence, when going from the non-dressed to the dressed potential, any changes that arise can be mapped directly to either the higher order properties of the Rydberg trapping potential or the influence of a spatially dependent light shift.

In Figures \ref{fig:dressedmf2}(a)-(b) the trapping potential of the dressed ground state atom is illustrated for the configuration $B=25\,$G, $G=2.5\,\mathrm{Tm}^{-1}$, $\omega_{ps}^{(0)}=2\pi\times 100\,$MHz, $\omega_{np}^{(0)}=2\pi\times 130\,$MHz, $\Delta_1=-2\pi\times 40\,$GHz, and $\Delta_2=-2\pi\times 1.5\,$MHz. In this strongly Ioffe field dominated case, the contribution $E_{40S}^{(2)}(\mathbf R)$ to the Rydberg trapping potential $E_{40S}(\mathbf R)$ is very strong, cf.\ (\ref{eq:dressealpha}). As a result, the Rydberg potential energy surface is extremely shallow and does not confine even a single center of mass state \cite{Mayle2009}. According to (\ref{eq:dressed2}), this strong deviation from the harmonic confinement of the ground state, $E_{5S}(\mathbf R)\propto \frac{1}{2}M\omega^2\rho^2$, is consequently mirrored in the dressed ground state potential: Along the diagonal ($X=Y$), where the effect of $E_{40S}^{(2)}(\mathbf R)$ is most pronounced, the trapping potential is gradually lowered compared to the harmonic confinement of the non-dressed ground state, cf.\ figure \ref{fig:dressedmf2}(b). Along the axes ($X=0$ or $Y=0$), on the other hand,  $E_\kappa^{(2)}(\mathbf R)$ vanishes and the non-dressed Rydberg and ground state energy surfaces coincide. As a consequence, the continuous azimuthal symmetry of the two-dimensional ground state trapping potential is reduced to a four-fold one, see figure \ref{fig:dressedmf2}(a). 

\begin{figure}
\centering
\includegraphics[width=8.6cm]{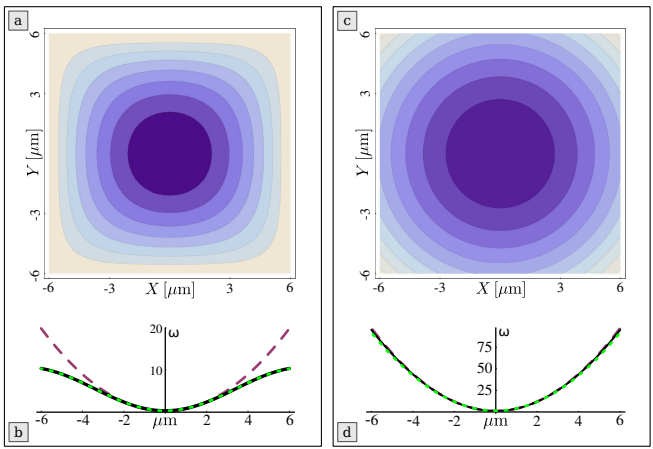}
\caption{(a) Contour plot of the dressed ground state trapping potential for $B=25\,$G, $G=2.5\,\mathrm{Tm}^{-1}$, $\omega_{ps}^{(0)}=2\pi\times 100\,$MHz, $\omega_{np}^{(0)}=2\pi\times 130\,$MHz, $\Delta_1=-2\pi\times 40\,$GHz, and $\Delta_2=-2\pi\times 1.5\,$MHz. (b) Cut along the diagonal $X=Y$ of the same surface (solid line); the short-dashed line which is on top of the solid black curve corresponds to the analytical solution (\ref{eq:dressed}) of the  simplified three level system. For comparison, the cut along the axis $X=0$ -- where $E_{40S}^{(2)}(\mathbf R)$ does not contribute -- is also illustrated, which corresponds to the trapping potential $E_{5S}(\mathbf R)$ of the ground state (dashed line). (c) and (d) Same as in subfigures (a) and (b), respectively, but for $B=1\,$G. In this case, all three curves coincide on the scale of figure (d). The energy scale of all subfigures is given by the ground state trap frequency $\omega=\sqrt{G^2/MB}$.}
\label{fig:dressedmf2}
\end{figure}

In addition to the full numerical solution, in figure \ref{fig:dressedmf2}(b) the results of the simplified three-level scheme according to (\ref{eq:dressed}) are illustrated as well (short-dashed line). Since in the Ioffe-field dominated regime the spatial variation of the quantization axis is minor, (\ref{eq:dressed}) agrees very well with the solution of the full 32-level problem (solid line). This allows us to recapitulate the above made observations on grounds  of the analytical expressions available within the reduced level scheme. For this reason, let us first consider the single photon Rabi frequencies as given by (\ref{eq:rabi1}-\ref{eq:rabi2}). Because of the strong Ioffe field, they experience only a weak spatial dependence and are therefore essentially defined by their values $\omega_{ps}^{(0)}$ and $\omega_{np}^{(0)}$ at the origin. As a consequence, also the light shifts $V_n$ and $V_s$ [cf.\  (\ref{eq:vn}-\ref{eq:vs})] as well as the effective two-photon Rabi frequency $\Omega$ [cf.\ (\ref{eq:omega})], can be approximated by their values at the origin. Hence, these quantities are not contributing to the particular shape of the dressed ground state energy surface. Omitting in this manner all contributions from (\ref{eq:dressed2}) that merely yield a constant energy offset, one arrives at 
\begin{equation}\label{eq:eminus}
 E_-(\mathbf R)=\tilde E_{5S}(\mathbf R)+\frac{\Omega^2}{4\Delta_2^2}E_{40S}^{(2)}
(\mathbf R)+\mathrm{const}.
\end{equation}
That is, the deviation of the dressed ground state surface from its non-dressed counterpart is given by 
\begin{equation}
 E_-(\mathbf R)-E_{5S}(\mathbf R)=\frac{\Omega^2}{4\Delta_2^2}E_{40S}^{(2)}(\mathbf R)\,.
\end{equation}
This complies with the observations made before: $E_\kappa^{(2)}(\mathbf{R})$ possesses the envisaged discrete azimuthal symmetry (${\mathbf C}_{4v}$) and contributes mostly close to the diagonals of the two-dimensional trapping surface while vanishing on the axes. Moreover, $E_{40S}^{(2)}(\mathbf R)<0$ which agrees with the lowering of the energy surface. Hence, the regime of strong Ioffe fields in combination with a strong laser coupling allows us to map the specific features of the Rydberg trapping potential onto the ground state.

In Figure \ref{fig:dressedmf2}(c)-(d) the same dressed trapping potential as before is illustrated, but now for $B=1\,$G. The reduction of the Ioffe field strength has basically two effects. First of all, considering the same spatial range as before the variation of the quantization axis is stronger. Consequently, the simplified three-level approach starts to slightly deviate from the exact solution, as can be observed in figure \ref{fig:dressedmf2}(d). Secondly, decreasing the Ioffe field influences the dressed potential by altering the Rydberg surface. For $B=1\,$G, $G=2.5\,\mathrm{Tm}^{-1}$, the Rydberg trapping potential is not quite as shallow as for $B=25\,G$, $G=2.5\,\mathrm{Tm}^{-1}$ and now supports a few confined center of mass states \cite{Mayle2009}. Consequently, the deviation between the Rydberg and the ground state surface is not as strong as in the previous case, resulting in a reduced lowering of the energy surface along the diagonal. Considering the two-dimensional trapping potential, the azimuthal symmetry is thus nearly recovered. In view of (\ref{eq:eminus}) this can be understood as follows. While the contribution $E_{40S}^{(2)}(\mathbf R)$ is identical for both cases (it depends only on the magnetic field gradient $G$ rather than on the Ioffe field strength $B$), the spatial dependence of $E_{5S}(\mathbf R)$ is stronger in the case of the weaker Ioffe field. Hence, for a decreasing Ioffe field the importance of $E_{40S}^{(2)}(\mathbf R)$ is diminished and the original behavior of the ground state trapping potential $E_{5S}(\mathbf R)$ is more and more recovered. Note that this does \emph{not} imply a smaller contribution of the Rydberg level to the dressed state. This regime is thus particularly useful if any change of the trapping surface due to the Rydberg dressing is not desirable but the admixture of the Rydberg character is still wanted. 

Regarding the magnetic field parameters, the previous example represents the intermediate regime between the Ioffe dominated and the gradient dominated case; the latter let us investigate in the following. To achieve a strong gradient Ioffe-Pritchard configuration, we further reduce the Ioffe field to $B=0.25\,$G and leave the magnetic field gradient $G=2.5\,\mathrm{Tm}^{-1}$ unchanged. An important aspect of the strong gradient regime is the contribution of $E_{40S}^{(2)}(\mathbf R)$ to the dressed ground state energy surface. Already in the case of figures \ref{fig:dressedmf2}(b)-(d) it was indicated that the influence of $E_{40S}^{(2)}(\mathbf R)$ is diminished if the gradient field becomes more important. Indeed, for the present field parameters the deviation of the Rydberg trapping potential from the ground state potential is minor and many center of mass states can be confined. Thus in the spatial domain we are considering, the continuous azimuthal symmetry of the ground state trapping potential is conserved and we present in figure \ref{fig:dressedmf2_2} only cuts along the diagonal of the dressed ground state energy surface. The parameters of the lasers are $\omega_{ps}^{(0)}=2\pi\times750\,$MHz, $\omega_{np}^{(0)}=2\pi\times 100\,$MHz, $\Delta_1=-2\pi\times 75\,$GHz, and $\Delta_2=-2\pi\times 5\,$MHz.

The dashed line in figure \ref{fig:dressedmf2_2}(a) represents the non-dressed ground state trapping potential, $E_{5S}(\mathbf R)$. As one can observe, the two-photon dressing (solid line) substantially alters this surface by significantly reducing the trap frequency, namely, from $2\pi\times638\,$Hz to $2\pi\times381\,$Hz. Although the simplified three-level system derived in Section \ref{sec:elimination} is not able to reproduce this result quantitatively, it nevertheless provides us a qualitative understanding of the underlying physics, as we shall demonstrate in the following. In the case of a strong gradient field, the light shift $V_s$ experienced by the ground state atom, (\ref{eq:vs}), shows a strong spatial dependence and therefore cannot be omitted in (\ref{eq:dressed2}). Specifically, it can be approximated for small center of mass coordinates by 
\begin{equation}
V_s\approx V_s^{(0)}\cdot\Big(1-\frac{1}{2}\frac{G^2\rho^2}{B^2}\Big)\,,
\end{equation}
where $V_s^{(0)}=-\frac{1}{4}\frac{|\omega_{ps}^{(0)}|^2}{\tilde E_p-\tilde E_s+\Delta_1}$ denotes the light shift at the origin. Except for constant contributions, the dressed ground state surface then reads
\begin{eqnarray}\label{eq:dressedintermediate}
 E_-(\mathbf R)&\propto& E_{5S}(\mathbf R)-\frac{1}{2}\frac{G^2\rho^2}{B^2}V_s^{(0)}\\
&=&\frac{1}{2}M(\omega^2-\frac{G^2}{M B^2}V_s^{(0)})\rho^2,
\end{eqnarray}
i.e., one encounters a reduced trap frequency $\tilde\omega^2\equiv \omega^2-\frac{G^2}{M B^2}V_s^{(0)}$. Note that the azimuthal symmetry of $E_-(\mathbf R)$ is conserved; hence the dressed trapping potentials experienced in this regime are qualitatively different from the one of figures \ref{fig:dressedmf2}(a)-(b). We stress that (\ref{eq:dressedintermediate}) only serves for our qualitative understanding of the underlying physics. In the given regime, it fails to quantitatively reproduce the dressed potentials. The actual spatial dependence of the light shift is illustrated as the short-dashed line in figure \ref{fig:dressedmf2_2}(b). It has been calculated by solving the full 32-level system but without the contribution of the Ioffe-Pritchard trapping potentials. The combination with the confinement $E_{5S}(\mathbf R)$ (dashed line) finally yields the surface of reduced trap frequency (solid line). 

The short-dashed line in figure \ref{fig:dressedmf2_2}(a) represents the trapping potential for $\omega_{np}^{(0)}=0$, i.e., in absence of the second laser that couples to the Rydberg state. Remarkably, turning off the second laser hardly changes the dressed potential. Hence, for the given example it is the interplay between the spatially varying quantization axis of the Ioffe-Pritchard field and the fixed polarization of the first laser that determines the spatially dependent light shift. As in the case of figures \ref{fig:dressedmf2}(c)-(d), this does not mean that the Rydberg state does not contribute to the dressed state. Hence, in the strong gradient regime we have two means to manipulate a ground state atom: With the first laser, one can alter the trapping potential experienced by the dressed atom and with the second laser we can in addition admix some Rydberg character to the atomic wave function.

\begin{figure}
\centering
\includegraphics[width=8.6cm]{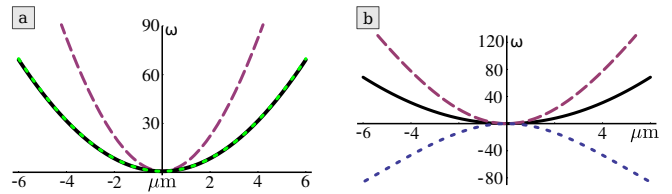}
\caption{(a) Cut along the diagonal of the dressed ground state trapping potential for $B=0.25\,$G, $G=2.5\,\mathrm{Tm}^{-1}$, $\omega_{ps}^{(0)}=2\pi\times750\,$MHz, $\omega_{np}^{(0)}=2\pi\times 100\,$MHz, $\Delta_1=-2\pi\times 75\,$GHz, and $\Delta_2=-2\pi\times 5\,$MHz. Note that in this strong gradient regime the trapping potential shows a continuous azimuthal symmetry, hence the corresponding contour plot is not provided. The trap frequency of the dressed surface (solid line) is greatly reduced compared to the trapping potential $E_{5S}(\mathbf R)$ of the ground state (dashed line). Turning off the second laser, i.e., setting $\omega_{np}^{(0)}=0$ hardly changes the shape of the potential surface (short-dashed line, on top of the black solid curve). (b) Spatially dependent light shift (short-dashed line) that in combination with the energy surface of the ground state (dashed line) leads to the trapping potential presented in subfigure (a) (solid line). The energy scale of all subfigures is given by the ground state trap frequency $\omega=\sqrt{G^2/MB}$.}
\label{fig:dressedmf2_2}
\end{figure}

The configuration leading to figure \ref{fig:dressedmf2_2} has one drawback: Since the influence of the spatial dependent light shift on the trapping potential strongly depends on the coupling strength to the intermediate state, the effective lifetime of the dressed state is restricted. The issue of the finite lifetime is discussed in more detail in section \ref{sec:exissues}. For now, let us remark that for the particular case of figure \ref{fig:dressedmf2_2} a lifetime of $\approx 1$ms can be achieved. Hence, the proposed scheme is suitable for scenarios where a short-term manipulation of the trapping potential is needed, e.g., for the modulation of the trap frequency on short timescales.

\subsection{Dressed Trapping Potentials of the $m_F=0$ State}
In the discussion above, we focused on the dressed ground state arising from the $m_F=2$ magnetic sublevel of the $5S_{1/2},F=2$ electronic state. Since ultracold samples of ground state atoms can nowadays routinely prepared and magnetically trapped in this state, this is a sensible choice. Nevertheless, also different magnetic sublevels merit a closer look. As an example, we consider in the following dressed states of the $m_F=0$ state. Note that the latter is untrapped in a pure Ioffe-Pritchard trap, i.e., without the coupling lasers. Therefore, one can expect that the influence of the specific features of the Rydberg trapping potential on the shape of the dressed surface is much more pronounced than in the case of the $m_F=2$ dressed state. Both examples that are presented in the following belong to the strong gradient regime where the simplified three-level scheme is not valid anymore and the full 32-level system must be considered.

In Figures \ref{fig:dressedmf0}(a)-(b) the trapping potential of the dressed $m_F=0$ ground state atom is illustrated for the configuration 
$B=1\,$G, $G=10\,\mathrm{Tm}^{-1}$, $\omega_{ps}^{(0)}=2\pi\times 100\,$MHz, $\omega_{np}^{(0)}=2\pi\times 35\,$MHz, $\Delta_1=-2\pi\times 14\,$GHz, and $\Delta_2=-2\pi\times 10\,$MHz. The first thing to note is that -- in contrast to the non-dressed $m_F=0$ state -- the atom experiences a confining potential that is due to the spatially dependent light shift of the off-resonant laser coupling. Moreover, Figure \ref{fig:dressedmf0}(a) reveals the four-fold symmetry known from the Rydberg trapping potential $E_{40S}(\mathbf R)$. Because the admixed Rydberg surface has not to compete against a strong magnetic confinement of the ground state according to $\boldsymbol{\mu}_F\cdot\mathbf{B(R)}$, the anti-trapping effect of $E_{40S}^{(2)}(\mathbf R)$ becomes particularly visible in the dressed potential of the $m_F=0$ state. In Figure \ref{fig:dressedmf0}(b) once more the cut along the diagonal of the dressed potential is illustrated (solid line). As expected, the admixture of the Rydberg surface eventually changes the character of the trapping potential from confining to de-confining when going to larger center of mass coordinates. However, for very large coordinates a weak confining behavior is recovered that can be explained as follows. For such large center of mass coordinates, the contribution $E_{40S}^{(2)}(\mathbf R)$ shifts the Rydberg state far off-resonant and thereby diminishes the contribution of the Rydberg level to the dressed state. The slightly confining character of the dressed potential in this regime is reminiscent of the spatially dependent light shift. Note that the azimuthally symmetric dressed potential arising in absence of the second laser (short-dashed line) coincides very well with the two-photon dressed potential along the axes (dashed line). Hence, the first laser can be used to trap and prepare the atoms in the $m_F=0$ ground state. By switching on the second laser, the Rydberg state gets admixed, resulting in the above described significant change of the trapping potential in the vicinity of the diagonals ($X=Y$). Overall, the influence of the Rydberg surface is much more distinct than in the case of the $m_F=2$ dressed states, cf.\ figure \ref{fig:dressedmf2}.

\begin{figure}
\centering
\includegraphics[width=8.6cm]{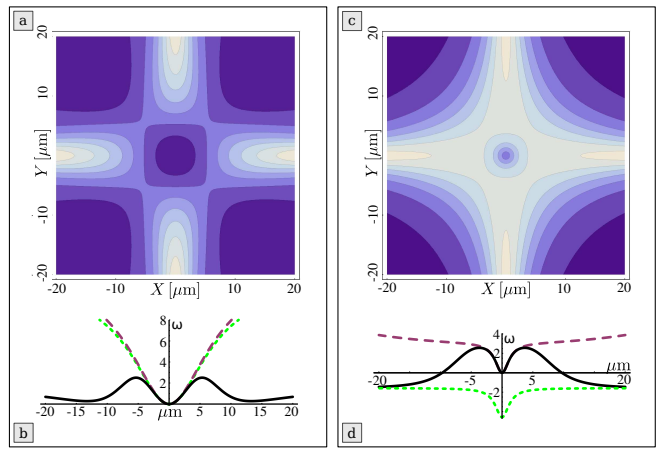}
\caption{(a) Contour plot of the dressed ground state trapping potential for the $5S_{1/2},m_F=0$ state. The parameters are $B=1\,$G, $G=10\,\mathrm{Tm}^{-1}$, $\omega_{ps}^{(0)}=2\pi\times 100\,$MHz, $\omega_{np}^{(0)}=2\pi\times 35\,$MHz, $\Delta_1=-2\pi\times 14\,$GHz, and $\Delta_2=-2\pi\times 10\,$MHz. (b) Cut along the diagonal $X=Y$ (solid line) and along the axis with $X=0$ (dashed line) of the same surface; the short-dashed line corresponds to the single-photon dressing, i.e., $\omega_{np}^{(0)}=0$. For comparison, all curves are offset to zero at the origin. (c) and (d) Same as in subfigures (a) and (b), respectively, but for $B=0.1\,$G, $G=10\,\mathrm{Tm}^{-1}$, $\omega_{ps}^{(0)}=2\pi\times 150\,$MHz, $\omega_{np}^{(0)}=2\pi\times 50\,$MHz, $\Delta_1=-2\pi\times 15\,$GHz, and $\Delta_2=-2\pi\times 10\,$MHz. In subfigure (d) we refrained from offsetting all curves to zero at the origin but rather applied a common offset such that the joint asymptote of the solid and dotted line becomes evident. Note that the detunings are defined in the same way as for the $5S_{1/2},m_F=2$ case. The energy scale in all subfigures is given in terms of the trap frequency. The latter has been gained by a harmonic fit around the origin, yielding $\omega=2\pi\times 25\,$Hz and $\omega=2\pi\times 187\,$Hz for the first and second configuration, respectively.}
\label{fig:dressedmf0}
\end{figure}

For comparison, we show in figures \ref{fig:dressedmf0}(c)-(d) the dressed trapping potentials of the $m_F=0$ state for a more dominant gradient field. The actual parameters are 
$B=0.1\,$G, $G=10\,\mathrm{Tm}^{-1}$, $\omega_{ps}^{(0)}=2\pi\times 150\,$MHz, $\omega_{np}^{(0)}=2\pi\times 50\,$MHz, $\Delta_1=-2\pi\times 15\,$GHz, and $\Delta_2=-2\pi\times 10\,$MHz. This configuration results in a much tighter confinement ($\omega=2\pi\times 187\,$Hz compared to $\omega=2\pi\times 25\,$Hz for the previous example) and a deeper trap along the diagonals. On the other hand, the revival of the weak trapping character as previously observed in figure \ref{fig:dressedmf0}(b) for large center of mass coordinates along the diagonal is lost since the light shift of the first laser already reached a constant asymptotic behavior in this regime. As discussed for the previous example, the contribution $E_{40S}^{(2)}(\mathbf R)$ shifts the Rydberg state far off-resonant and thereby diminishes the influence of the second laser on the excitation dynamics. Consequently, for large center of mass coordinates close to the diagonal the dressed surface approaches the asymptotic of the single-photon dressing of the first laser [short-dashed line in figure \ref{fig:dressedmf0}(d)]. In contrast, along the axes (dashed line) the potential does not reach a constant asymptote but maintains a weak confining behavior. The latter is due to the admixture of predominantly $m_j=1/2$ Rydberg states: Since we assumed the two-photon transition to be blue-detuned, the magnetic field interaction $\propto m_j|\mathbf{B(R)}|$ pushes the $m_j=1/2$ Rydberg state closer to resonance and in the same manner repels its $m_j=-1/2$ counterpart. Hence, the dressed state shows a stronger admixture of trapped than anti-trapped Rydberg states, giving rise to a confining energy surface. We remark that not only the Rydberg state contributes to the dressed state. In fact, the first laser slightly mixes the $5S_{1/2},F=2,m_F=0$ state with $m_F\neq0$ states of the same hyperfine level. However, the $m_F<0$ states are admixed in the same degree as their $m_F>0$ counterparts such that their confining and de-confining characters cancel.

We remark that, as in the case of figure \ref{fig:dressedmf2_2}, the effective lifetime of the dressed $m_F=0$ states is restricted to a few ms due to the coupling to the intermediate state. The latter is required for the confining light-shift potential of the otherwise untrapped $m_F=0$ state.

\subsection{Experimental Issues}
\label{sec:exissues}
Let us finish by commenting on the experimental feasibility of the above discussed scheme. The proposed dressed states possess a finite lifetime due to the spontaneous decay of the Rydberg state. The associated effective lifetime can be estimated by 
\begin{equation}\label{eq:efflife}
\tau=\frac{\tau_n}{|c_n|^{2}}\,,
\end{equation}
$c_n$ being the admixture coefficient of the Rydberg state; within the simplified three-level scheme it evaluates to $|c_{n,\mathrm{3l}}|^2=[\Omega/2\tilde\Delta_2]^{2}$ with $\tilde\Delta_2=\Delta_2+V_n-V_s+E_\mathrm{hfs}$. $\tau_n$ denotes the radiative lifetime of the $nS_{1/2}$ Rydberg atom and can be parameterized as $\tau_n=\tau'(n-\delta)^\gamma$ where one finds
$\tau'=1.43$ ns and $\gamma=2.94$ for $l=0$,
$\tau'=2.76$ ns and $\gamma=3.02$ for $l=1$, and
$\tau'=2.09$ ns and $\gamma=2.85$ for $l=2$ \cite{PhysRevA.65.031401}.
For the $40S_{1/2}$ Rydberg state, this yields $\tau_{40}=58\,\mu$s. In table \ref{tab:efflife}, the effective lifetimes for the examples presented in this work are tabulated. Because the Rydberg state is only weakly admixed ($|c_n|^2<10^{-2}$ for all examples), effective lifetimes greater than ten milliseconds are obtained. 
Besides the finite lifetime of the Rydberg state, one needs in addition to account for the decay of the intermediate $5P_{3/2}$ level that possesses a much shorter radiative lifetime of $\tau_p=26\,$ns. The resulting effective lifetimes together with the coupling coefficients are provided in table \ref{tab:efflife} as well. As it turns out, the decay of the intermediate state constitutes the dominating loss channel, allowing for lifetimes $\gtrsim 1\,$ms.

The resulting lifetimes need to be compared with the timescales emerging in actual experiments. In case of figure \ref{fig:dressedmf2}(a), it is desirable to map the dressed potential by the external motion. Given the trap frequency of $\omega=2\pi\times63.8\,$Hz, this yields a timescale of about 15 ms which is of the same order of magnitude as the effective lifetime. For the remaining examples addressed in this work, the effective lifetime is not quite sufficient to cover the timescale of the external motion. Hence, in these cases the proposed scheme is more suitable whenever only a short term manipulation of the trapping potential is required. Finally, we remark that the effective lifetime can be further prolonged by coupling to Rydberg states with a higher principle quantum number $n$ and by substituting the intermediate $5P_{3/2}$ state by a more long-lived one such as the $6P_{3/2}$ state.

\begin{table}
\caption{Effective lifetimes of the dressed states considered in this section. The lifetimes are determined according to (\ref{eq:efflife}) using the Rydberg admixture coefficient $c_n$ of the full 32-level system. When applicable, the admixture coefficient $c_{n,\mathrm{3l}}$ of the simplified three-level system is provided for comparison. In addition, the admixture coefficient $c_p$ of the intermediate state and the resulting lifetime are given. \label{tab:efflife}}
\begin{ruledtabular}
 \begin{tabular}{l c c c c c}
  Configuration & $|c_{n,\mathrm{3l}}|^2$&$|c_n|^2$&$\tau_n$(ms)&$|c_p|^2$&$\tau_p$(ms)\\
\hline
 figure \ref{fig:dressedmf2}(a)&$0.004\,186$&$0.004\,134$&$14.0$&$1.83\times10^{-6}$&$14.2$\\
 figure \ref{fig:dressedmf2}(c)&$0.004\,533$&$0.004\,472$&$13.0$&$1.84\times10^{-6}$&$14.1$\\
 figure \ref{fig:dressedmf2_2}(a)&$-$&$0.001\,433$&$40.5$&$25.2\times10^{-6}$&$1.03$\\
 figure \ref{fig:dressedmf0}(a)&$-$&$0.000\,040$&$1457$&$12.8\times10^{-6}$&$2.03$\\
 figure \ref{fig:dressedmf0}(c)&$-$&$0.000\,154$&$377$&$25.2\times10^{-6}$&$1.03$\\
\end{tabular}
\end{ruledtabular}
\end{table}

Similar to the effective lifetime, the van der Waals interaction of two Rydberg atoms is suppressed by $|c_n|^4$. The latter interaction results in an energy shift $\Delta_\mathrm{vdW}$ that depends on the interparticle distance and effectively alters the detuning of the two-photon transition. In order to avoid any such effects, $\Delta_\mathrm{vdW}$ should be well below the the excitation detuning $\Delta_2$. Taking $c_n=0.1$ and $\Delta_\mathrm{vdW}<2\pi\times 0.1$\,MHz as an (quite restrictive) example yields a minimum interparticle distance of $\approx 1\,\mu$m.

\section{Summary}
In the present work, we investigated a magnetically trapped rubidium atom that is coupled to its $nS$ Rydberg state via a two-photon laser transition. We studied the off-resonant case where the ground state atom becomes \emph{dressed} by the Rydberg state and vice versa. By this procedure, the peculiar properties of Rydberg atoms become accessible also for ground state atoms. In particular, we explored how the trapping potential experienced by a ground state atom in a magnetic Ioffe-Pritchard trap can be manipulated by means of such an off-resonant laser coupling. It is demonstrated that in the limit of a strong offset field the four-fold azimuthal symmetry, which is inherent for the trapping potential of the Rydberg atom, is mirrored in the dressed ground state trapping potential. In this regime, a simplified three-level scheme is derived that facilitates the interpretation of the observed results. In the opposite regime of a strong gradient, the delicate interplay between the spatially varying quantization axis of the Ioffe-Pritchard field and the fixed polarizations of the laser transitions greatly influences the actual shape of the dressed trapping potentials. In this manner, the trapping potentials of ground state atoms can be manipulated substantially.

\begin{acknowledgments}
This work was supported by the German Research Foundation (DFG) 
within the framework of the
Excellence Initiative through the Heidelberg Graduate School of
Fundamental Physics (GSC~129/1).
M.M.\ acknowledges financial support from the Landesgraduiertenf\"orderung
Baden-W\"urttemberg.
Financial support by the DFG through the grant Schm 885/10-3
is gratefully acknowledged.
\end{acknowledgments}

\appendix
\section*{Appendix}
\setcounter{section}{1}
In this appendix the two-level Hamiltonian (\ref{eq:h2l}) is derived. Within the rotating wave approximation \cite{scully97} our initial, time-independent Hamiltonian in the simplified three-level scheme reads
\begin{equation}\label{eq:dressH0V}
 \mathcal H_\mathrm{rwa}=
\underbrace{
\left(\begin{array}{ccc}
 \Delta_1+\tilde{E}_p&0&0\\
  0&\Delta_2+\tilde{E}_n+E_\mathrm{hfs}&0\\
  0&0&\tilde{E}_s
\end{array}\right)}_{\equiv \mathcal H_0}+
\underbrace{
\left(\begin{array}{ccc}
  0&\frac{\omega_{pn}}{2}&\frac{\omega_{ps}}{2}\\
  \frac{\omega_{np}}{2}&0&0\\
  \frac{\omega_{sp}}{2}&0&0
 \end{array}\right)}_{\equiv \mathcal V}.
\end{equation}
Here, $\tilde E_n\equiv \frac{1}{2}|\mathbf{B(R)}|+C\cdot G^2X^2Y^2$, $\tilde E_s\equiv \frac{1}{2}|\mathbf{B(R)}|$, and $\tilde E_p\equiv |\mathbf{B(R)}|$ are the trapping potentials of the individual energy levels and $E_\mathrm{hfs}$ includes the energy shift due to the hyperfine splitting of the Rydberg state as well as the Zeeman shift of the nuclear spin. The laser detunings are defined by $\Delta_1=E_p^{el}-E_s^{el}-\omega_1$ and $\Delta_2=E_n^{el}-E_s^{el}-\omega_1-\omega_2$.
The second term, $\mathcal V$, represents a perturbation that couples the ``model space'' $\{n,s\}$ consisting of the $nS_{1/2}$ Rydberg and the $5S_{1/2},F=m_F=2$ ground state via the single-photon Rabi frequencies $\omega_{ps}$ and $\omega_{np}$, cf.\ (\ref{eq:rabi1}-\ref{eq:rabi2}), to the one-dimensional orthogonal subspace $\{p\equiv5P_{3/2},F=m_F=3\}$ of the intermediate level.

We are considering the regime where the intermediate level $p$ is only weakly coupled to both the ground state $s$ and the Rydberg level $n$. Such a scenario allows us to adiabatically eliminate the intermediate state $p$ from the excitation dynamics. Specifically, if $|\Delta_1|\gg\omega_{ps}$ and $|\Delta_1-\Delta_2|\gg\omega_{np}$, quasidegenerate van Vleck perturbation theory provides us a unitary transformation $e^{-G}$ that block diagonalizes $\mathcal H_\mathrm{rwa}$ \cite{shavitt:5711}. In this manner, the subspace $\{p\}$ is decoupled from the dynamics of the model space $\{n,s\}$, yielding an effective Hamiltonian $H_{2l}=H_0+W$ for the latter. Our goal is to determine the unitary transformation $e^{-G}$ and hence the effective interaction $W$ within the model space $\{n,s\}$.

The formalism to calculate $G$ and accordingly $W$ is derived in \cite{shavitt:5711}, which we briefly summarize for our system in the following.
As shown in (\ref{eq:dressH0V}), the Hamiltonian $H_\mathrm{rwa}$ can be divided into a zero-order part $H_0$ and a perturbation $V$ with zero-order eigenfunctions $H_0|t\rangle=\varepsilon_t|t\rangle$. The set of eigensolutions of $H_0$ can be partitioned into two subsets $\{t,u,\dots\}=\{\alpha,\beta,\dots\}\cup\{i,j,\dots\}$, defining the model space $\{\alpha,\beta,\dots\}=\{n,s\}$ and its orthogonal complement $\{i,j,\dots\}=\{p\}$. The projection operator into the model space and its orthogonal complement read $P=|n\rangle\langle n|+|s\rangle\langle s|$ and $Q=1-P=|p\rangle\langle p|$, respectively. Any operator $A$ can be partitioned into a block diagonal part $A_D$ and a block off-diagonal part $A_X$, $A=A_D+A_X$.
In our case, the perturbation $V$ only possesses block off-diagonal matrix elements, i.e., $V_D=0$ and we find $H_D=H_0$ and $H_X=V_X$.

Within the canonical form of van Vleck perturbation theory, we require $G=G_X$, i.e., $G_D=0$ for the operator determining the unitary transformation $e^{-G}$. Moreover, $G$ is an anti-Hermitian operator, $G=-G^\dagger$. It is defined order by order via
\begin{eqnarray}
{}  [H_0,G^{(1)}]&=&-V_X\,,\label{eq:dressGn1}\\
{}  [H_0,G^{(2)}]&=&-[V_D,G^{(1)}]=0\,.\label{eq:dressGn}
\end{eqnarray}
For higher orders, see \cite{shavitt:5711}; the zeroth order contribution vanishes, i.e., $G^{(0)}=0$. The order-by-order computation of the effective interaction $W$ follows as
\begin{eqnarray}\label{eq:dressWn}
  W^{(1)}&=&V_D=0\,,\\
  W^{(2)}&=&\frac{1}{2}[V_X,G^{(1)}]\,.
\end{eqnarray}
Explicit equations for the $G^{(n)}$ can be gained from (\ref{eq:dressGn1}-\ref{eq:dressGn}) using the resolvent formalism. 
In first order, one finds the matrix representation
\begin{eqnarray}\label{eq:dressGja}
 \mathcal G_{j\alpha}^{(1)}&=&\frac{\mathcal V_{j\alpha}}{\varepsilon_\alpha-\varepsilon_j}\,.
\end{eqnarray}
Note that per definition there are no block diagonal contributions, i.e., $\mathcal G_{\alpha\beta}^{(n)}=0$. The second order matrix elements of the effective interaction $W$ correspondingly read
\begin{eqnarray}\label{eq:dressWba}
\mathcal W_{\beta\alpha}^{(2)}&=&\frac{1}{2}\sum_i\mathcal V_{\beta i}\mathcal V_{i\alpha}\Big(\frac{1}{\varepsilon_\alpha-\varepsilon_i}+\frac{1}{\varepsilon_\beta-\varepsilon_i}\Big)\\
&=&\frac{1}{8}\omega_{\beta p}\omega_{p \alpha}\Big(\frac{1}{\varepsilon_\alpha-\varepsilon_p}+\frac{1}{\varepsilon_\beta-\varepsilon_p}\Big)\,.
\end{eqnarray}
Hence, the above described procedure provides us an effective two-level system whose excitation dynamics are determined up to second order by the Hamiltonian
\begin{equation}
 \mathcal H_{2l}=\mathcal H_0+\mathcal W^{(2)}=
\left(\begin{array}{ccc}
 \Delta_2+\tilde{E}_n+E_\mathrm{hfs}+V_n& \Omega/2\\
 \Omega^*/2&\tilde{E}_s+V_s
\end{array}\right)
\end{equation}
where the effective interaction between the ground- and Rydberg state is given by the two-photon Rabi frequency 
\begin{equation}
 \Omega=\frac{\omega_{ps}\omega_{np}}{4}
\left[\frac{1}{\tilde{E}_s-\tilde{E}_p-\Delta_1}+\frac{1}{\tilde{E}_n-\tilde{E}_p+\Delta_2-\Delta_1+E_\mathrm{hfs}}\right].
\end{equation}
The contributions
\begin{eqnarray}
 V_s&=-\frac{1}{4}\frac{|\omega_{ps}|^2}{\tilde{E}_p-\tilde{E}_s+\Delta_1}\,,\\
 V_n&=-\frac{1}{4}\frac{|\omega_{np}|^2}{\tilde{E}_p-\tilde{E}_n+\Delta_1-\Delta_2-E_\mathrm{hfs}}\,,
\end{eqnarray}
are the light shifts of the ground and Rydberg state, respectively, which stem from the off-resonant laser dressing of the individual single-photon transitions \cite{cohen85}.

\end{document}